# Yosys+nextpnr: an Open Source Framework from Verilog to Bitstream for Commercial FPGAs

David Shah[*†], Eddie Hung[‡*], Clifford Wolf[*], Serge Bazanski[*], Dan Gisselquist[*] and Miodrag Milanović[*]
[*]Symbiotic EDA; Vienna, Austria; {`david`,`clifford`}`@symbioticeda.com`
[†]Dept. of Electrical and Electronic Engineering; Imperial College London, UK
[‡]Dept. of Electrical and Computer Engineering; University of British Columbia, Canada; `eddieh@ece.ubc.ca`

*Abstract*—This paper introduces a fully free and open source software (FOSS) architecture-neutral FPGA framework comprising of Yosys for Verilog synthesis, and nextpnr for placement, routing, and bitstream generation. Currently, this flow supports two commercially available FPGA families, Lattice iCE40 (up to 8K logic elements) and Lattice ECP5 (up to 85K elements) and has been hardware-proven for custom-computing machines including a low-power neural-network accelerator and an OpenRISC system-on-chip capable of booting Linux. Both Yosys and nextpnr have been engineered in a highly flexible manner to support many of the features present in modern FPGAs by separating architecture-specific details from the common mapping algorithms. This framework is demonstrated on a longest-path case study to find an atypical single source-sink path occupying up to 45% of all on-chip wiring.

## I. INTRODUCTION

In many ways, Field-Programmable Gate Arrays can be likened to an "Etch-A-Sketch" plotting device — a blank canvas for near-limitless creativity that can be enjoyed (in reconfigurable fashion) by young and old, but one that remains difficult to truly master. Far from being a toy though, FPGA technology continues to be extensively studied in academia and to gain traction in industry, including most recently, large scale deployments in cloud datacenters.

Continuing with the Etch-A-Sketch analogy where access to the substrate is controlled solely through the two knobs on its front face, access to FPGA silicon is equivalently available only through the use of closed-source Computer-Aided Design (CAD) tools provided exclusively by the FPGA vendor. This sole entry point belies a major disconnect between the two worlds of academia and industry: the inability for certain innovations made in the former to be physically realised on the latter, on real silicon. The Yosys+nextpnr framework described herein represents a first step to breaking out of this walled garden for those that wish to experiment with the creation of custom-computing machinery outside.

Typically, academic researchers have turned to existing open-source frameworks such as Verilog-to-Routing (VTR) [1] to investigate their hypotheses. Long regarded as the de-facto FPGA research framework, the VTR project allows questions to be asked of (a) the underlying FPGA architecture (for example, the optimal number of LUT inputs) as well as (b) the best CAD algorithms for use in mapping to such architectures. In order to support the ability to answer questions of type (a), necessarily VTR must target theoretical architectures (perhaps one modelled on commercial architectures [2]) that can be procedurally generated from a number of architectural parameters. Currently, there is no native publicly-available support for commercial devices within, with prior work finding that extending VTR to support non-ideal, real-world FPGA architectures can be a difficult affair [3] and that conclusions made on its theoretical architectures can be misleading when applied onto commercial devices [4].

In contrast, Yosys+nextpnr does not focus on exploring theoretical architectures but instead has been carefully engineered to natively target commercial-off-the-shelf (COTS) FPGAs, including all their imperfections and real-world considerations. Figure 1 shows the positioning of Yosys+nextpnr relative to VTR and vendor tools.

## II. RELATED WORK

*Synthesis:* Open-source Verilog synthesis is supported by Odin II, which is tightly integrated into the Verilog-to-Routing [1] project. Odin II accepts a synthesisable subset of Verilog as Yosys does, supports hard-IP such as the block-RAMs, carry-chains, and multipliers present in VTR architectures, but can only produce BLIF netlists primarily used in research. In contrast, Yosys not only supports the BLIF format (demonstrated on VTR in [3]) but also supports EDIF and Verilog output formats interoperability with commercial tools, and allows flexible coarse-grained technology mapping (and inference) as described in the following section. Lastly, Titan [2] is an extension of the VTR flow that replaces its Odin II front-end with Intel's (closed-source) vendor tools, enabling downstream experiments to be made with larger and more complex benchmarks not otherwise supported.

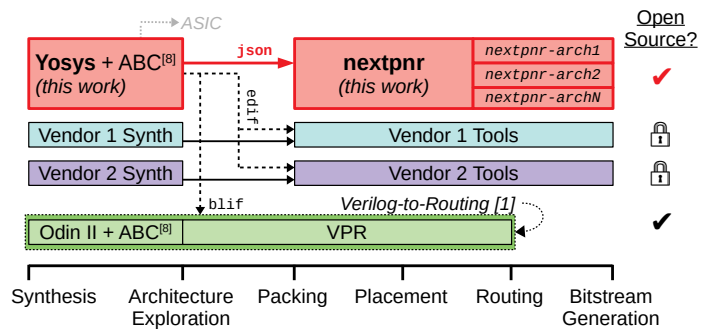

Fig. 1. Comparison of nextpnr against VPR and vendor FPGA tools.

*Place-and-route:* Versatile Place and Route (VPR) has been a mainstay of academic research (and even a number of commercial ventures!) since its inception over 20 years ago, and is packaged as part of the VTR project [1]. VTR architectures are described theoretically using the XML format, detailing the (proportional) makeup and layout of soft and hard blocks on the targeted FPGA, using architectural parameters such as number of LUT inputs, as well as their local and global routing connectivities. Importantly, and in contrast to nextpnr and to vendor tools, VPR's ability to target procedurally generated architectures enables new architectures to be designed and limit studies to be made — for example, to measure architectural efficiency by determining the smallest FPGA that a benchmark can fit on, or to quantify the performance of routing algorithms by determining the minimum routing channel width necessary — studies that cannot be made when supporting only a fixed set of discretely-sized devices.

*Commercial architecture support:* A number of projects exist for open-source support of real-world devices. These include VTR-to-Bitstream [3] that overrides VTR's procedural capabilities with actual device data to support a Xilinx Virtex-6 device, as well as frameworks such as RapidSmith [5] that provide a sandbox, inclusive of simple packing, placement and routing algorithms, for experimenting with multiple Xilinx device families. However, such projects are not solely the domain of academics as evidenced by RapidWright [6] from Xilinx Research Labs that provides an "escape-hatch" into Vivado.

## III. YOSYS – VERILOG SYNTHESIS

Yosys [7] is an open-source framework for Verilog synthesis and verification. It supports all commonly-used synthesisable features of Verilog-2005, and can target both FPGAs and ASICs. Yosys uses ABC [8] for logic optimisation and LUT/cell mapping; combined with custom coarse-grained optimisations and dedicated passes for inferring and mapping block- and distributed-RAM, flip-flops and arithmetic structures.

A typical FPGA flow, after logic elaboration, would perform some coarse-grain optimisations and map the results to a set of generic hard-logic cells. Generic passes are then used to infer block-RAM, flip-flops supporting clock-enables and set-resets, arithmetic logic and more, followed by architecture-specific technology mapping. Any remaining coarse-grain cells are converted to gates by Yosys and then mapped to LUTs by ABC. Further architecture-specific rules then map generic LUT and flip-flop cells to the target device's primitives.

Coarse- and fine-grained cells are mapped (even recursively) according to a Verilog description. As an example, Figure 2 shows how a generic 8-input LUT can be transformed into two LUT7s plus a dedicated multiplexer, and from there onto a total of four LUT6s and three muxes when synthesising for Xilinx 7-series. This capability gives Yosys a high degree of flexibility, while reducing the effort for targeting new architectures. Yosys currently supports synthesis for the Xilinx 7-series, Lattice

```verilog
// Apply these mapping rules to Yosys' generic LUT cells
module \$lut (A, Y);
  parameter WIDTH = 0; // Number of LUT inputs
  parameter LUT = 0;   // LUT mask contents
  input [WIDTH-1:0] A; // LUT input signals
  output Y;            // LUT output signal
  generate
    wire T0, T1;
    if (WIDTH == 8) begin
      // Map a generic 8-input LUT to two generic 7-inputs
      \$lut #(.WIDTH(7), .LUT(LUT[127:0]) fpga_lut_0 (
        .O(T0), .A(A[6:0]));
      \$lut #(.WIDTH(7), .LUT(LUT[255:128])) fpga_lut_1 (
        .O(T1), .A(A[6:0]));
      // ... plus a target-specific mux primitive
      MUXF8 fpga_mux_0 (.O(Y), .I0(T0), .I1(T1), .S(A[7]));
    end else if (WIDTH == 7) begin
      // Decompose a generic 7-input LUT into two target-
      specific 6-input LUTs, plus another specialised mux
      LUT6 #(.INIT(LUT[63:0])) fpga_lut_0 (.O(T0),
        .I0(A[0]), .I1(A[1]), .I2(A[2]),
        .I3(A[3]), .I4(A[4]), .I5(A[5]));
      LUT6 #(.INIT(LUT[127:64])) fpga_lut_1 (.O(T1),
        .I0(A[0]), .I1(A[1]), .I2(A[2]),
        .I3(A[3]), .I4(A[4]), .I5(A[5]));
      MUXF7 fpga_mux_0 (.O(Y), .I0(T0), .I1(T1), .S(A[6]));
    end else begin
      // Internal marker to indicate no mapping otherwise
      wire _TECHMAP_FAIL_ = 1;
    end
  endgenerate
endmodule
```

Fig. 2. Example of Yosys' flexible technology-mapping capabilities for recursively decomposing 7- and 8-input LUTs into 6-input ones with muxes.

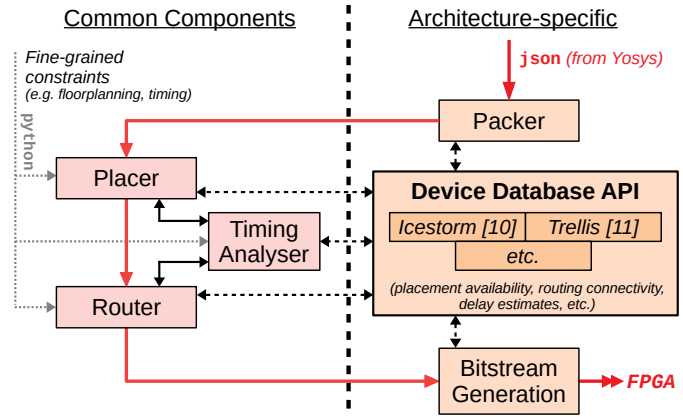

Fig. 3. Breakdown of nextpnr's shared (left) versus architecture-specific (right) components; at its core is the "Device Database API" that answers basic queries, such as availability and validity of the placer's decisions.

iCE40, Lattice ECP5, and Silego GreenPAK4; as well as experimental support for Intel, Gowin and Anlogic families.

## IV. NEXTPNR – PACK, PLACE, ROUTE & BITSTREAM-GEN.

nextpnr is an open-source, timing-driven, place-and-route framework targeting real-world FPGA silicon supporting Linux, Windows and macOS platforms. Unlike many existing tools which describe an architecture using a flat file format such as XML; an architecture in nextpnr is an implementation of an Application Program Interface (API).

This gives nextpnr the flexibility to support the irregularities and intricacies of modern commercial FPGAs. An architecture can provide its own custom packer; functionality to check validity and availability of placement sites, routing switches and wires, its own timing model, and other custom passes as needed — such as a dedicated global network routing pass, or bitstream generation. Architectures must also provide their own implementation of the device database.

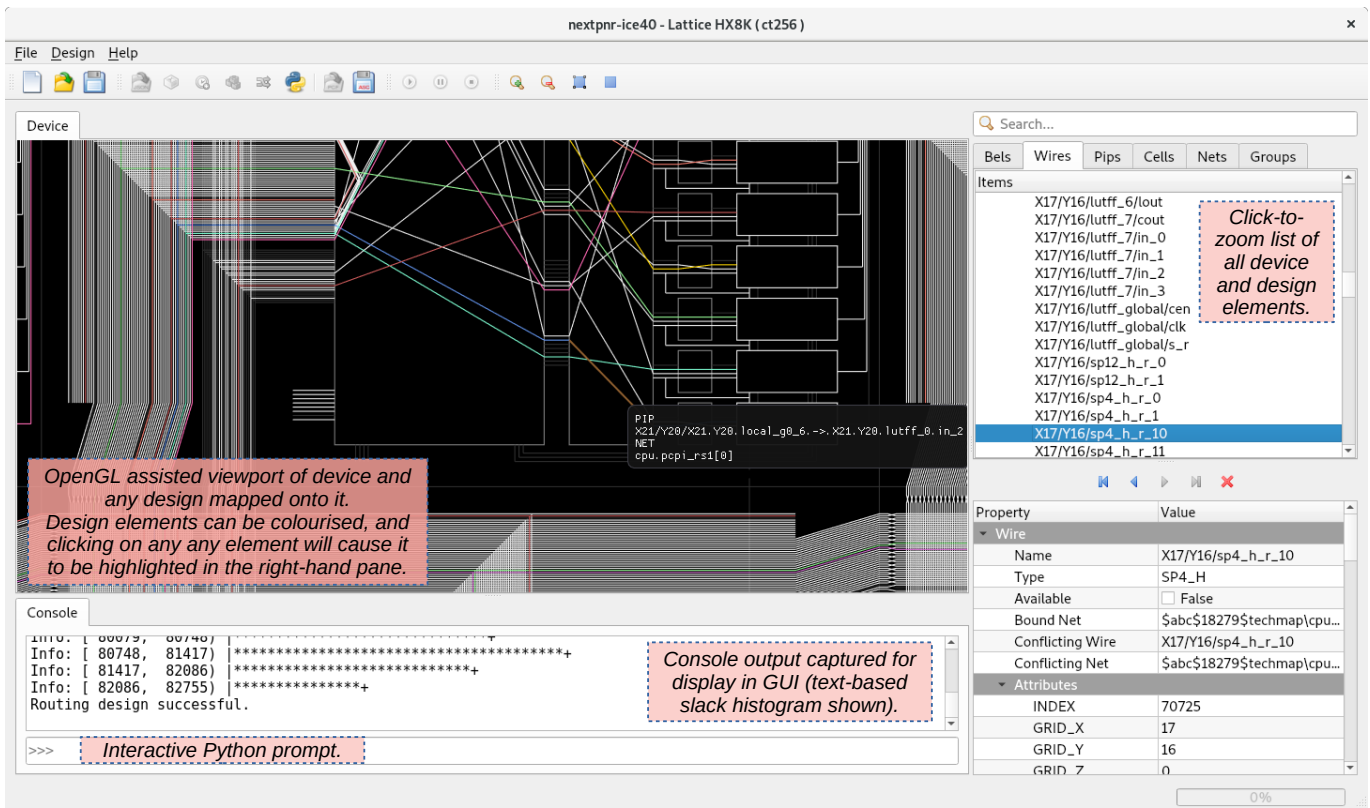

Fig. 4. nextpnr Graphical User Interface.

This flexibility is implemented without detriment to performance by each architecture providing its own set of source files that implement a source-compatible (as opposed to binary-compatible) interface, and building a separate executable for each architecture. Doing so also avoids the overhead of virtual (polymorphic) functions for frequently called code and allows compile-time optimisations such as inlining.

A breakdown of nextpnr's internals is shown in Figure 3, differentiating between its common and architecture-specific components. nextpnr uses a JSON-based design interchange format for netlists, a format that supports cell names, inout ports, attributes and parameters unavailable in BLIF whilst being easier to process than formats such as EDIF or structural Verilog. Packing is an architecture-specific stage responsible for transforming primitives from the input netlist into larger cells, for example, in the iCE40 target LUTs and their fanout flip-flops are merged into a LUTFF cell, though the API allows use of other granularities. Two timing-driven placers are included, traditional simulated annealing and an analytic placer based on HeAP[9] supporting relative and complex floorplanning constraints. Routing is also timing-driven and currently based on a naïve iterative rip-up and re-route A* algorithm, followed by bitstream generation. Furthermore, nextpnr provides a modern graphical user interface that allows navigation through any implemented design as well as the underlying architecture, as shown in Figure 4.

The iCE40 and ECP5 flows employ information from the open-source Project Icestorm [10] and Project Trellis [11] respectively to provide architecture-specific data (e.g. location and connectivity of logic and wiring resources, as well as cell and interconnect timing) and to generate bitstreams. This makes them fully independent from the vendor FPGA tools. Besides basic soft-logic, carry-chains, and block-RAM, a large amount of hard-IP is also supported across both flows including PLLs, DSPs, DDR inputs/outputs, constant-current LED drivers, hard PWM generators, and gigabit transceivers.

*Device database deduplication*: Since nextpnr allows implementation details to be deferred to individual architectures through a source-level interface, this enables custom features such as database deduplication. Essentially, this is a one-off compile-time 'compression' process that takes advantage of an FPGA's inherent symmetry and redundancy to reduce the disk and memory footprint of its database, thus improving cache-efficiency and runtime.

Currently, deduplication is implemented for the ECP5 family with devices of up to 85K logic elements and 4M wires and 28M routing switches. The uncompressed database for the largest device is 1GB; after deduplication this database shrinks to 38MB. Moreover, this scales with device size improves considerably, the deduplicated database for the 85K device is just 13% larger than that for the 25K device.

*Python API*: To facilitate development of external plug-ins and allow custom design flows for real-world devices, nextpnr's internal API is exposed to Python through use of the Boost::Python library. Although almost every feature of the C++ API is exposed to Python, a shim exists to reverse some performance optimisations in order to make the Python API considerably easier to use.

As well as being able to replace the entire nextpnr flow with a custom Python script, scripts can also be run before or after any stage of the default flow (for example, between packing and placement). This makes Python useful as a way to specify constraints, for floorplanning, to perform custom placement or routing for a subset of cells or nets, or to analyse/export any in-progress design. It is worth noting that the Python and GUI are optional functionality, opening the door to running Yosys+nextpnr on embedded platforms such as FPGA-SoCs.

## V. Example Application: Longest Path Routing

As a brief and teasing example of the freedom available with an open-source framework targeting physical devices, we sought to demonstrate the antithesis of typical CAD tools — rather than searching for the shortest routing path between two pins, what about finding the longest? The longest path problem for directed cyclic graphs is known to be an NP-hard problem; however, similar to the traditional objective of seeking the routing solution with the globally minimum critical-path, we also apply a heuristic to find a path as long as possible. One utility of constructing such paths may be for online test.

Our experiment consisted of mapping an input Verilog file with a single input, a single output (constrained to pins 20 and 12 respectively), and an assign statement connecting the two onto a Lattice iCE40 UltraPlus UP5K device (SG48 package) containing 125K wires (nodes in a directed graph) and 1.3M programmable routing switches (edges). The input is synthesised using Yosys (commit `5387ccb`) and then implemented using a modified version of nextpnr (branched from commit `c46a22c`) to replace its shortest-path router with the longest simple path heuristic (simple refers to each wire being visited at most once) from [12] that adopts a successive-relaxation approach. In total only 87 lines of nextpnr code were added and experiments were conducted on a mid-range 14nm AMD 2400G desktop CPU running Ubuntu 18.04.

TABLE I
Physically measured path delay $T_{delay}$ (in microseconds) and used fraction of all reachable routing wires $F_{wires}$.

| Objective | | ~0s | <10s | <30s | <300s | <3000s |
|---|---|---|---|---|---|---|
| Shortest | $T_{delay}$ | ≤0.01 | | | | |
| | $F_{wires}$ | 0.0004 | | | | |
| Longest | $T_{delay}$ | - | 18.47 | 21.21 | 25.21 | 25.42 |
| | $F_{wires}$ | - | 0.358 | 0.391 | 0.445 | 0.451 |

The results of Table I compares the delay ($T_{delay}$) between the existing shortest-path algorithm and the new longest-path heuristic; since Yosys+nextpnr can target physical devices, these represent real-world delay values measured using a 100 MS/s logic analyser. Also shown is the fraction of all reachable routing wires and LUT route-throughs that were used in this path — where reachability is defined by the corresponding node's existence in the strongly connected component extracted from the routing graph in which all IO blocks except for the design's input and output ports have been removed. This statistic reflects the upper bound of wires that can appear on any path, though there is no guarantee that a (Hamiltonian) solution doing so can exist. Our example application is available at https://github.com/eddiehung/nextpnr-lsp for reproduction.

## VI. Conclusion

Yosys+nextpnr is a framework that implements Verilog designs onto COTS FPGA devices. Currently, two Lattice families are supported, and work is underway to target those from other vendors. For researchers, by innovating within this framework users will gain the ability to easily evaluate on a broad range of real-silicon devices with different architectures and process nodes. For industry, the aim of Yosys+nextpnr is to lower the barrier for deploying CAD innovations onto existing devices, and to accelerate time-to-market and reduce risk for supporting novel new FPGA architectures. Yosys and nextpnr are both available under a permissive (thus facilitating commercial use) open-source license from http://github.com/YosysHQ and between them, are the ongoing product of over 100 contributors.

We do not regard nor intend to position Yosys+nextpnr as a threat to existing commercial flows and do not expect to be competitive with them. Instead, we envision a *symbiotic* arrangement where both tool flows can co-exist: certified vendor flows will continue to be used for production, mission-critical designs; but for more experimental or incremental tasks and for unorthodox custom-computing applications, the fine-grained control available through our framework may be more suitable. We are particularly enthused by how this vision aligns with that of Xilinx's own RapidWright [6] project.